# RedisGraph GraphBLAS Enabled Graph Database

Pieter Cailliau[1], Tim Davis[2], Vijay Gadepally[3], Jeremy Kepner[3],
Roi Lipman[1], Jeffrey Lovitz[1], Keren Ouaknine[1]
[1]RedisLabs, [2]Texas A&M, [3]MIT

## Abstract
RedisGraph is a Redis module developed by Redis Labs to add graph database functionality to the Redis database. RedisGraph represents connected data as adjacency matrices. By representing the data as sparse matrices and employing the power of GraphBLAS (a highly optimized library for sparse matrix operations), RedisGraph delivers a fast and efficient way to store, manage and process graphs. Initial benchmarks indicate that RedisGraph is significantly faster than comparable graph databases.

## I. Introduction

One of the most pressing issues with big data is the ability to store, manage and examine hundreds of millions of data points that exist beyond traditional boundaries, determine their relationship with each other and deliver new insights to existing context. The speed with which distributed big data can be analyzed is often critical in making sense of the various datasets.

RedisGraph leverages complex and dynamic relationships in highly connected data to deliver new insights and intelligence across a variety of different use cases, including real-time recommendation engines, personalization, fraud detection, cyber security, master data management, social networking, 360-degree customer view and many more.

RedisGraph implements an enhanced matrix traversal methodology representing connected data as sparse adjacency matrices and adopts a standardized engine from GraphBLAS.org, that uses linear algebra and compressed matrix representation to overcome the performance and scale challenges. RedisGraph simplifies the traversal of highly connected, variable data to answer complex questions and deliver contextual insights.

## II. RedisGraph Architecture

RedisGraph leverages the GraphBLAS standard to achieve high performance for its graph operations. The GraphBLAS exploit the duality between matrices and graphs to provide a small number of highly optimized operations that enable a wide range of graph analytics [1–3]. These mathematics have been developed into a C standard library and in implemented the SuiteSparse GraphBLAS library [4–6]

RedisGraph exposes an API with the graph query language Cypher [7]. A Cypher query gets translated by RedisGraph into a query execution plan of a.o. graph traversals, that get translated into linear algebraic operations on sparse matrices leveraging GraphBLAS.

Redis is a single-threaded process by default. Having all data within a single shard (a horizontal partition of data in a database) avoids network overhead between shards. RedisGraph is bound to the single thread of Redis to support all incoming queries and includes a threadpool that takes a configurable number of threads at the module's loading time to handle higher throughput. Each graph query is received by the main Redis thread, but calculated in one of the threads of the threadpool. This allows reads to scale and handle large throughput easily. Each query, at any given moment, only runs in one thread.

This differs from other graph database implementations, which execute each query on all available cores of the machine. We believe our approach is more suitable for real-time real-world use cases where high throughput and low latency under concurrent operations are more important than processing a single serialized request at a time.

## III. Benchmarking

In the graph database space, there are multiple benchmarking tools available. The most comprehensive one is LDBC graphalytics [8], but for the initial general availability release of RedisGraph, we opted for a simpler benchmark released by TigerGraph [9]. The benchmark evaluated leading

This material is based in part upon work supported by National Science Foundation grants DMS-1312831 and CCF-1533644.

graph databases like TigerGraph, Neo4J [10], Amazon Neptune [11], JanusGraph [12], and ArangoDB [13], and published the average execution time and overall running time of all queries on all platforms. The TigerGraph benchmark covers the query response time for k-hop neighborhood count, for k=1,2,3 and 6.

The benchmark system used for RedisGraph measurements and in the literature consisted of an AWS r4.8xlarge with 32 vCPUs, 244 GB RAM, 10 Gigabit network, and EBS-Only SSD storage. The graph datasets were drawn from Twitter (41.6M vertices and 1.47B edges) and the Graph500 (2.4M vertices and 67M edges) benchmark data generator [14].

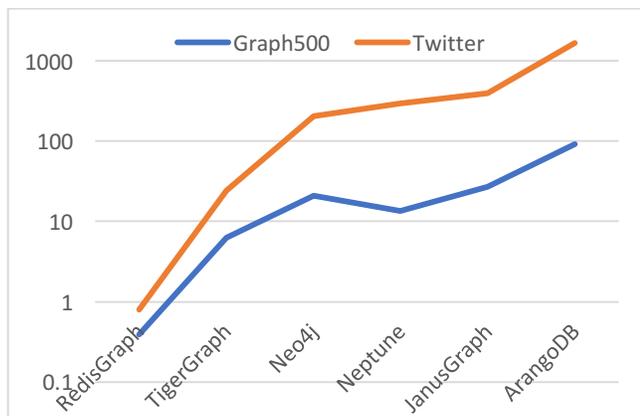

**Fig. 1**. Average response time in msec for 1-hop queries on Graph500 and Twitter graphs for RedisGraph and various databases [9].

The single request benchmark tests reported here are based on 300 seeds for the one and two-hop queries, and on 10 seeds for the three and six-hop queries. These seeds are executed sequentially on all graph databases. The measured average response time results for RedisGraph are shown in Fig. 1. The performance of other graph databases taken from the literatures are also shown for comparison.

### IV. Conclusions & Future Work

RedisGraph outperforms Neo4j, Neptune, JanusGraph, and ArangoDB on a single request response time with improvements 36 to 15,000 times faster. RedisGraph achieved 2X and 0.8X faster single request response times compared to TigerGraph, which uses all 32 cores to process the single request compared to RedisGraph which uses only a single core. It is also important to note that none of the queries timed out on the large data set, and none of them created out of memory exceptions.

During these tests, profiling RedisGraph found additional opportunities for enhancement: aggregations and large result sets, incorporation of enhanced GraphBLAS, cypher clauses/functionality to support more diverse queries, integration of graph visualization software, utilization of custom GraphBLAS hardware [15], and further benchmarking on LDBC and GraphChallenge [16, 17]. Future work will focus on evaluating these enhancements.